# A Study of Bianchi Type-I Cosmological Model With Cosmological Constant

Mukesh Sharma and Sonia Sharma[*]
*BAM Khalsa College, Garhshankar, India*
*Rayat Polytechnic College, SBS Nagar, India*

Einstein's field equations with variable cosmological constant is considered in the presence of a perfect fluid for a Bianchi type-I universe by assuming that the cosmological term is proportional to the square of the Hubble parameter. The variation law for vacuum density was recently proposed by many researchers on the basis of quantum field estimation in a curved expanding background. The cosmological term tends asymptotically to a genuine cosmological constant and the model tends to a de-Sitter universe. Here obtained some new results by using a slightly different method from that of other researchers obtained the result that the present universe is accelerating with a large fraction of cosmological density in the form of a cosmological term.

## 1. Introduction

One of the most remarkable observational discoveries in recent times is the cosmological constant problem, which is very interesting to all researchers. The cosmological constant $\Lambda$ was originally given by Einstein in his field equations. In an evolving universe, it appears natural to look at this constant as a function of time. The $\Lambda$ term arises naturally in general relativistic quantum field theory, where it is interpreted as the energy density of the vacuum [1-4]. Some authors [5-9] argued for the dependence of $\Lambda$. Cosmological models with variable G and $\Lambda$ have been studied by a number of researchers [10-17] for a homogeneous and isotropic FRW line element. Also, Bianchi type-I models are studied by using variable G and $\Lambda$ [18-24]. Schutzhold [25,26] recently proposed that the vacuum energy density is proportional to the Hubble parameter, which leads to vacuum energy density decaying as $\Lambda \approx m^3 H$, where $m \approx 150 MeV$ is the energy scale of the chiral phase transmission of QCD. Also, Borges and Carnerio [27] have considered an isotropic and homogeneous flat space filled with matter and a cosmological term proportional to H obeying the equation of state of the vacuum. Recently Tiwari and Divya Singh [28] investigated the anisotropic Bianchi type-I model with a varying $\Lambda$ term. Tiwari and Sonia [29] investigated the non-existence of shear in Bianchi type-III string cosmological models with bulk viscosity and time-dependent $\Lambda$. Also Tiwari and Sonia [30] investigated the Bianchi type-I string cosmological model with bulk viscosity and time-dependent $\Lambda$ term. For studying the possible effects of anisotropy in the early universe on present day observations many researchers [31-35] have investigated Bianchi type-I models from a different point of view.

In the present paper, we investigate the homogeneous anisotropic Bianchi type-I space time with variable $\Lambda$ containing matter in the form of a perfect fluid. We obtain the solution of the Einstein field equations assuming that cosmological term is proportional to Hubble parameter for stiff matter.

## 2. Metric and Field Equations

The line element for spatially homogeneous and anisotropic Bianchi type-I space-time is described by

$$ds^2 = -dt^2 + A^2 dx^2 + B^2 dy^2 + C^2 dz^2 \qquad (1)$$

Where, A, B, C are functions of *t* only.

We assume that cosmic matter is represented by the energy momentum tensor of a perfect fluid

$$T_{ij} = (\rho + p)v_i v_j + p g_{ij} \qquad (2)$$

Where, ρ and p are energy density and thermo-dynamic pressure, and $v_i$ is the four velocity vector of the fluid satisfying the relation $v_i v^i = -1$. We assume that the matter content obeys an equation of state,

_________
[*]soniamathematics@yahoo.co.in

$$p = \omega\rho, 0 \leq \omega \leq 1 \quad (3)$$

The Einstein's equations with varying $\Lambda$ in suitable units are

$$R_{ij} - \frac{1}{2}Rg_{ij} = -T_{ij} + \Lambda g_{ij} \quad (4)$$

Spatial volume V as an average scale factor of the model (1) may be defined as

$$V = R^3 = ABC \quad (5)$$

Hubble parameter H in anisotropic models may be defined as

$$H = \frac{\dot{R}}{R} = \frac{1}{3}\left(\frac{\dot{A}}{A} + \frac{\dot{B}}{B} + \frac{\dot{C}}{C}\right) \quad (6)$$

Where a dot stands for ordinary time derivative of the concerned quantity

$$H = \frac{1}{3}(H_1 + H_2 + H_3) \quad (7)$$

Where, $H_1 = \frac{\dot{A}}{A}, H_2 = \frac{\dot{B}}{B}$ and $H_3 = \frac{\dot{C}}{C}$ are directional Hubble factors in the x, y, and z directions, respectively.

For the metric (1) and energy moment tensor (2) in the commoving system of coordinates, the field equation (4) yields

$$\frac{\ddot{B}}{B} + \frac{\ddot{C}}{C} + \frac{\dot{B}\dot{C}}{BC} = -p + \Lambda \quad (8)$$

$$\frac{\ddot{A}}{A} + \frac{\ddot{C}}{C} + \frac{\dot{A}\dot{C}}{AC} = -p + \Lambda \quad (9)$$

$$\frac{\ddot{A}}{A} + \frac{\ddot{B}}{B} + \frac{\dot{A}\dot{B}}{AB} = -p + \Lambda \quad (10)$$

$$\frac{\dot{A}\dot{B}}{AB} + \frac{\dot{B}\dot{C}}{BC} + \frac{\dot{A}\dot{C}}{AC} = \rho + \Lambda \quad (11)$$

In view of the vanishing divergence of the Einstein tensor, we have

$$\left[\dot{\rho} + (\rho + p)\left(\frac{\dot{A}}{A} + \frac{\dot{B}}{B} + \frac{\dot{C}}{C}\right)\right] + \dot{\Lambda} = 0 \quad (12)$$

The non-vanishing component of shear tensor $\sigma_{ij}$ defined by $\sigma_{ij} = u_{i;j} + u_{j;i} - \frac{2}{3}g_{ij}u^k_k$ are obtained as

$$\sigma^1_1 = \frac{4}{3}\frac{\dot{A}}{A} - \frac{2}{3}\left(\frac{\dot{B}}{B} + \frac{\dot{C}}{C}\right) \quad (13)$$

$$\sigma^2_2 = \frac{4}{3}\frac{\dot{B}}{B} - \frac{2}{3}\left(\frac{\dot{C}}{C} + \frac{\dot{A}}{A}\right) \quad (14)$$

$$\sigma^3_3 = \frac{4}{3}\frac{\dot{C}}{C} - \frac{2}{3}\left(\frac{\dot{A}}{A} + \frac{\dot{B}}{B}\right) \quad (15)$$

Thus the shear scalar $\sigma$ is given by

$$\sigma^2 = \frac{1}{3}\left(\frac{\dot{A}^2}{A^2} + \frac{\dot{B}^2}{B^2} + \frac{\dot{C}^2}{C^2} - \frac{\dot{A}\dot{B}}{AB} - \frac{\dot{B}\dot{C}}{BC} - \frac{\dot{A}\dot{C}}{AC}\right) \quad (16)$$

$$\therefore \frac{\dot{\sigma}}{\sigma} = -\left(\frac{\dot{A}}{A} + \frac{\dot{B}}{B} + \frac{\dot{C}}{C}\right) = -3H \quad (17)$$

The Einstein's field equations (8)-(11) in terms of Hubble parameter H, shear scalar $\sigma$ and declaration parameter q can be written as

$$H^2(2q-1) - \sigma^2 = p - \Lambda \quad (18)$$

$$3H^2 - \sigma^2 = \rho + \Lambda \quad (19)$$

Where

$$q = -1 - \frac{\dot{H}}{H^2} = -\frac{R\ddot{R}}{\dot{R}^2} \quad (20)$$

From Eqns. (8), (9) and (10) and integrating, we get

$$\frac{\dot{A}}{A} - \frac{\dot{B}}{B} = \frac{k_1}{ABC} = \frac{k_1}{R^3} \quad (21)$$

$$\frac{\dot{A}}{A} - \frac{\dot{C}}{C} = \frac{k_2}{ABC} = \frac{k_2}{R^3} \quad (22)$$

$$\frac{\dot{B}}{B} - \frac{\dot{C}}{C} = \frac{k_3}{ABC} = \frac{k_3}{R^3} \quad (23)$$

Where, $k_1$, $k_2$, $k_3$ are constants of integration. We now assume energy conservation equation $T_{ij}=0$ yields

$$\dot{\rho} + \rho(1+\omega)\left(\frac{\dot{A}}{A} + \frac{\dot{B}}{B} + \frac{\dot{C}}{C}\right) = 0 \qquad (24)$$

Using Eqns. (5) & (24), we get

$$\rho = \frac{k_4}{R^{3(1+\omega)}} \qquad (25)$$

Where, $k_4$ is the constant of integration.

Again integrating Eqns. (21)-(23), we get

$$\frac{A}{B} = m_1 \exp\left(k_1 \int \frac{1}{R^3} dt\right) \qquad (26)$$

$$\frac{A}{C} = m_2 \exp\left(k_2 \int \frac{1}{R^3} dt\right) \qquad (27)$$

$$\frac{B}{C} = m_3 \exp\left(k_3 \int \frac{1}{R^3} dt\right) \qquad (28)$$

Where, $m_1, m_2, m_3$ are integration constants.

From Eqn. (20), we obtain

$$\frac{3\sigma^2}{\theta^2} = 1 - \frac{3\rho}{\theta^2} - \frac{3\Lambda}{\theta^2} \qquad (29)$$

$implying\ \Lambda \geq 0, 0 < \frac{\sigma^2}{\theta^2} \leq \frac{1}{3}, 0 < \frac{\rho}{\theta^2} < \frac{1}{3}.$

Thus the presence of positive $\Lambda$ lowers the upper limit of anisotropy whereas a negative value of $\Lambda$ gives more room for anisotropy.

Eqn. (29) can be written as

$$\frac{3\sigma^2}{3H^2} = 1 - \frac{\rho}{3H^2} - \frac{\Lambda}{3H^2} = 1 - \frac{\rho}{\rho_c} - \frac{\rho_v}{\rho_c} \qquad (30)$$

Where, $\rho_c = 3H^2$ is the critical density and $\rho_v = \Lambda$ is the vacuum density.

$$\frac{d\theta}{dt} = -\frac{3}{2}(\rho + p) - 3\sigma^2 \qquad (31)$$

Showing that the rate of volume expansion decreases during time evolution and the presence of positive $\Lambda$ slows down the rate of decrease whereas a negative $\Lambda$ would promote it. From Eqns. (19) and (20),

$$\Lambda = (2-q)H^2 - \frac{(1-\omega)\rho}{2} \qquad (32)$$

This implies: $\Lambda \leq 0$ for $q \geq 2$.

The system of Eqn. (3), and Eqns. (8)-(11) are five independent equations in six unknowns A, B, C, $\rho$, p and $\Lambda$. Therefore, one extra condition is needed to solve the system completely. For this we take the cosmological term proportional to the Hubble parameter, since many authors considered it as $\Lambda$ decay. Schutzhold [26] consider variation law for vacuum density, Borges and Carnerio [27], R. K. Tiwari and Divya Singh [28], Tiwari and Sonia [29,30] have considered a cosmological term proportional to H. Thus we take the decaying vacuum energy density

$$\Lambda = \beta H^2 \qquad (33)$$

Where, $\beta$ is the positive constant.

Let $\Omega = \Lambda/\rho$ be the ratio between the vacuum and matter densities. From Eqns. (19) and (33), we get

$$\beta = \frac{3\Omega}{1+\Omega}\left(1 - \frac{\sigma^2}{27\theta^2}\right) \qquad (34)$$

Thus the value of $\beta$ in an anisotropic background is smaller in comparison to its value in isotropic.

For stiff fluid ($\omega = 1$), Eqns. (18), (19) and (33) lead to a differential equation

$$\dot{H} + (3-\beta)H^2 = 0 \qquad (35)$$

Integrating, we get

$$R = \left[(3-\beta)(c_1 t + c_2)\right]^{\frac{1}{3-\beta}} \qquad (36)$$

Where, $c_1$ and $c_2$ are the constant of integration.

$$H = \frac{\dot{R}}{R} = c_1 \left[(3-\beta)(c_1 t + c_2)\right]^{-1} \qquad (37)$$

$$A = \left[(3-\beta)(c_1 t + c_2)\right]^{\frac{1}{3-\beta}} \exp\left[\frac{2k_1 + k_2}{6\{(3-\beta)(c_1 t + c_2)\}^{\frac{3}{3-\beta}}}\right]$$

$$B = \left[(3-\beta)(c_1 t + c_2)\right]^{\frac{1}{3-\beta}} \exp\left[\frac{k_2 - k_1}{3\{(3-\beta)(c_1 t + c_2)\}^{\frac{3}{3-\beta}}}\right]$$

$$C = \left[(3-\beta)(c_1 t + c_2)\right]^{\frac{1}{3-\beta}} \exp\left[\frac{2k_2 - k_1}{2\{(3-\beta)(c_1 t + c_2)\}^{\frac{3}{3-\beta}}}\right]$$

Therefore, the metric (1) reduces to

$$ds^2 = -dt^2 + \left[(3-\beta)(c_1 t + c_2)\right]^{\frac{2}{3-\beta}} \left\{\exp\left[\frac{2k_1 + k_2}{3\{(3-\beta)(c_1 t + c_2)\}^{\frac{3}{3-\beta}}}\right]dx^2 + \exp\left[\frac{2(k_2 - k_1)}{3\{(3-\beta)(c_1 t + c_2)\}^{\frac{3}{3-\beta}}}\right]dy^2 + \exp\left[\frac{2k_2 - k_1}{\{(3-\beta)(c_1 t + c_2)\}^{\frac{3}{3-\beta}}}\right]dz^2\right\}$$

For this model, the matter density $\rho$, pressure p, cosmological term $\Lambda$, shear scalar $\sigma$ and expansion scalar $\theta$ are given by

$$\rho = p = k_4 \{(3-\beta)(c_1 t + c_2)\}^{\frac{-6}{3-\beta}}$$

$$\Lambda = \beta c_1^2 \{(3-\beta)(c_1 t + c_2)\}^{-2}$$

$$\theta = \frac{H}{3} = \frac{c_1}{3}\{(3-\beta)(c_1 t + c_2)\}^{-1}$$

$$\sigma = (c_1 t + c_2)^{\frac{-3c_1^2}{3-\beta}} c_3$$

The ratio between the vacuum and matter densities is given by

$$\Omega = \frac{\Lambda}{\rho} = \frac{\beta c_1^2}{k_4}\{(3-\beta)(c_1 t + c_2)\}^{\frac{2\beta}{3-\beta}}$$

The deceleration parameter q for this model is

$$q = 2 - \beta$$

The vacuum energy density $\rho_v$ and the critical density $\rho_c$ are given by

$$\rho_v = \beta c_1^2 \{(3-\beta)(c_1 t + c_2)\}^{-1}$$

$$\rho_c = 3c_1^2 \{(3-\beta)(c_1 t + c_2)\}^{-2}$$

Spatial volume

$$V = R^3 = \{(3-\beta)(c_1 t + c_2)\}^{\frac{3}{3-\beta}}$$

The spatial volume V is zero at $t=-c_2/c_1$ and expansion scalar is infinite at $t=-c_2/c_1$. It shows that the Universe starts evolving with zero volume with an infinite rate of expansion. The scale factor R is also zero at $t=-c_2/c_1$, which means that during initial age the space time exhibits a point type singularity. At $t=-c_2/c_1$, $\rho\to\infty$, $\sigma\to\infty$. As time increases, the scale factor R and spatial volume V increases, but the expansion scalar decreases i.e., the rate of expansion slows down. When time $t\to\infty$, then $R\to\infty$, $V\to\infty$, $\Lambda\to\infty$ and $\rho$, $\rho_c$, $\rho_v$ all tends to zero. Therefore, model gives an empty universe for $t\to\infty$. This result is in agreement with observations obtained by many astronomers [12,28,30,36]

In summary, we have investigated the Bianchi type-I cosmological model containing a stiff fluid with cosmological term $\Lambda=\beta H^2$. It is found that the deceleration parameter q for the model is 2 at $\beta=0$, it is zero at $\beta=2$ and decreases as $\beta$ increases. The cosmological term $\Lambda$, being very large at the initial stage, later relaxes to a genuine cosmological constant, which is in accordance with the recent observations [34-38]. The model asymptotically tends to the de-Sitter universe.